\documentclass[dvips]{arxstspdf}
\usepackage{flushend}
\usepackage{stfloats}

\volume{26}
\issue{2}
\pubyear{2011}
\firstpage{161}
\lastpage{161}
\doi{10.1214/11-STS359}

\begin{document}
\begin{frontmatter}

\title{Introduction}

\end{frontmatter}

The Statistics Consortium at the University of Maryland, College
Park, hosted a two-day workshop on Bayesian Methods that
Frequentists Should Know during April 30--May 1, 2008.  The event
was co-sponsored by the Institute of Mathematical Statistics
(IMS), Office of Research and Methodology, National Center for
Health Statistics, Survey Research Methods Section (SRMS) of the
American Statistical Association, and Washington Statistical
Society. The workshop was intended to bring out the positive
features of Bayesian statistics in solving real-life
problems, including complex problems in sample surveys and
production of high-quality official statistics.

The workshop consisted of six invited sessions, plus a poster
session held in the evening of April 30 where ten posters were
displayed and discussed in an informal setting. The average
attendance in the sessions was about 100.  The invited sessions
and the poster session together covered a wide range of topics,
including Bayesian methods in public
policy, missing data problems, objective prior selection, small
area estimation, sample surveys, relationship between parametric
bootstrap and Bayesian methods, and accurate approximation to
posterior densities. In keeping with the title of the workshop, the
audience included many attendees who did not classify themselves as
Bayesian statisticians, and the speakers took pains to explain the
attractiveness of Bayesian methods to non-Bayesians.

Shortly after the workshop, we discussed the possibility of
publishing a special issue of \textit{Statistical Science} with David
Madigan, the Executive Editor of \textit{Statistical Science}.  We
agreed that a special issue containing a few review papers around
the main theme of the workshop along with formal discussions and
rejoinders would be of great interest to \textit{Statistical Science}
readers. The intent was not to publish a workshop proceedings but
to provide an overview of materials presented in the workshop from
the perspective of a few experts in the field.

We believe that the five authoritative review papers contained in
this issue, along with their discussions and rejoinders, provide an
excellent overview of the assessment of the current state of the
Bayesian methodology. This issue will certainly be a valuable
resource for researchers, and will be of interest also to a general
statistical audience. We would like to thank all the authors,
discussants and referees for their hard work in meeting the high
standard of the journal. Finally, we would like to thank David
Madigan for his encouragement and patience during the entire
editorial process.%\vspace*{12pt}

\begin{flushright}
P. Lahiri and Eric Slud\\
University of Maryland, College Park\\
Guest Editors
\end{flushright}

\end{document}